\documentclass[twocolumn, prb,showpacs,superscriptaddress]{revtex4-1}	
\usepackage{amsmath}
\usepackage{graphicx}
\usepackage{hyperref}
\DeclareMathOperator{\tr}{Tr}

\newcommand{\exponential}[1]{\mathrm{e}^{#1}}

\begin{document}
\title{Low-temperature evolution of the spectral weight of a spin-up carrier
 moving in a ferromagnetic background}

\author{Mirko M. M\" oller}
\affiliation{\!Department \!of \!Physics and Astronomy, \!University of\!
  British Columbia, \!Vancouver, British \!Columbia,\! Canada,\! V6T \!1Z1}

\author{Mona Berciu}
\affiliation{\!Department \!of \!Physics and Astronomy, \!University of\!
  British Columbia, \!Vancouver, British \!Columbia,\! Canada,\! V6T \!1Z1}
\affiliation{\!Quantum Matter \!Institute, \!University of British Columbia,
  \!Vancouver, British \!Columbia, \!Canada, \!V6T \!1Z4} 

\date{\today}
\begin{abstract}
We derive the lowest-temperature
correction to the self-energy of a spin-up particle injected in a
ferromagnetic background. The background is modeled with both
Heisenberg and Ising Hamiltonians so that differences due to gapless
vs. gapped magnons  can be understood. Beside the expected thermal
broadening of the quasiparticle peak as it becomes a resonance
inside a continuum, we also find that spectral weight is
transferred to regions lying outside this continuum. We explain the
origin of this spectral weight transfer and its low-temperature
evolution. 
\end{abstract}
\maketitle
\section{Introduction}

The problem of understanding the behavior of a carrier doped into a
magnetically ordered insulator is relevant for the study of many
materials. Multiple variations are possible: the carrier may enter
into the same band that gives rise to the magnetic order, or, as if
often the case, may be hosted in a different band. The background
might have antiferromagnetic (AFM) order (parent
cuprates\cite{cuprates} being the most famous example), or 
ferromagnetic (FM) order (in ferromagnetic chalcogenides like EuO) or
more complicated forms of magnetic order, such as FM layers 
that are layered antiferomagnetically in manganite
perovskites,\cite{Colossal-Magneto-Resistance, Nolting2} zig-zag order
in iridates,\cite{iridates} etc. Finally, the magnetic order may
be present in the undoped compound (all examples listed above)
or may arise as a result of doping, like in diluted magnetic
semiconductors such as Ga$_{1-x}$Mn$_x$As,\cite{GaMnAs} or heavy fermion materials
like CeSi$_x$.\cite{CeSix, Nolting2} Understanding the properties of
such materials has direct
technological implications since many of them are 
candidates for new spintronic and magnetoelectric
devices.\cite{spintronics_review}

The degree of difficulty in solving such problems varies
widely. The most difficult problems are those with AFM backgrounds,
because of their inherent complexity due to the presence of
quantum spin fluctuations -- this explains why what happens when one
hole is doped into an AFM cuprate layer is still being debated.\cite{Bayo} 

In contrast, FM backgrounds are exactly solvable, especially at $T=0$.  On the
other hand, unlike in the AFM case, here the spectrum of 
the carrier  has a striking dependence on its
spin direction. If the carrier is injected with its
spin oriented parallel to the local moments, no spin-flip excitations are possible and
the carrier moves freely. Its spectrum is identical to that of a free
carrier up to an energy shift due to the $zz$ component of the
magnetic exchange. If the carrier is injected with
its spin anti-parallel to the local moments, the formation of a
dressed quasi-particle, a so-called spin-polaron, is possible. This is
a state where the carrier continuously emits and re-absorbs a magnon
while flipping its spin from up to down, in a coherent fashion. There
are also states where the carrier has spin-up and the magnon is
present (as required by conservation of the $z$-component of the total
spin) but not bound to the carrier, giving rise to a continuum of
incoherent states distinct from the spin-polaron discrete state.

The fact that only one magnon can be emitted by a spin-down carrier
injected in a $T=0$ FM background (assuming the carrier has
spin-$\frac{1}{2}$, which we  do here) explains why
there exists an exactly solvable solution for such problems. The
solution was first given by Shastry and Mattis\cite{Shastry} and has
recently been generalized to more complex lattices.\cite{Berciu-sp}
Furthermore, exact analytical derivations of the eigenstates and
eigenenergies were recently presented by Henning et al.\cite{Nolting} and
by Nakano et al. \cite{Nakano} for Hamiltonians describing such
problems.

As far as we know, the only other exact solution for a
generalization of this simplest case is for two carriers injected in
the FM background,\cite{Mirko} because the number of possible
additional magnons is still very small, resulting in a solvable
few-body problem. Dealing with finite carrier concentrations which can
induce finite concentrations of magnons requires the use of
approximations,\cite{Nolting-MCDA-RKKY} except in the very
trivial case when all carriers have spin-up.

Here we consider another and in real life more interesting generalization, namely that of
studying the spectrum of a spin-up carrier injected in a FM background
at finite $T$.  An exact solution is no longer
possible since one needs to consider states with arbitrary numbers of
magnons when performing temperature averages. A natural approach for
low-$T$ is to consider states with a small number of magnons;
this is what we do here. As a result, the solution we propose
becomes asymptotically exact in the limit of very low temperatures,
where ``low'' means well-below the Curie critical temperature $T_C$ of
the FM background.

As  mentioned, a spin-up carrier  has a very
simple spectrum at $T=0$, mirroring that of the free carrier, with a single
eigenstate for a given momentum. At $T\ne 0$ thermally
activated magnons are present in the system and the carrier can now
flip its spin by absorbing one of them. Interaction with even one such
magnon takes the problem in the Hilbert subspace appropriate for the
$T=0$ spin-down carrier, which has a very different spectrum. As
a result, we expect that spectral weight is transferred from the
spin-up quasiparticle peak to energies in the spectrum of
the spin-polaron, as $T$ increases. How exactly does this occur at very
low $T$, and what happens to the infinitely-lived discrete state that
was the only feature in the spectrum at $T=0$, is the topic of this work.

Furthermore, we consider two types of  exchange
between the local moments, namely  Heisenberg
exchange and Ising exchange (in both cases, the characteristic
energy scale is  $J$). For the latter the magnon spectrum is
gapped, whereas for the former the magnon spectrum is gapless. This
allows us to contrast the two 
cases to understand the relevance of the magnon's spectrum on the
evolution of the up carrier's spectral function with $T$.

Finite temperature studies have been previously carried out by Nolting
et al.\cite{Nolting2} for the Kondo lattice model (KLM), which is also
often referred to as the s-f model. This model accounts for the
kinetic energy of the carrier as described by a tight-binding
model with an energy scale $t$, and for the exchange between the local
moments and the carrier, described by a Heisenberg exchange
with a coupling $J_0$.  Unlike the models we consider, KLM does not
include the  exchange $J$ between local moments; this is
one key difference between our work and theirs. The second is the
approach employed. While, as mentioned, we use a low-$T$ expansion to
calculate the propagator, Nolting et al. proposed an ansatz for the
self-energy chosen so as to reproduce asymptotic limits
where an exact solution is available, specifically the $T=0$ solution
mentioned above and the case of finite $T$ but zero bandwidth,
$t=0$.\cite{Nolting-zero-bandwith} (This approach was later
generalized to finite carrier concentrations as
well.\cite{Nolting-MCDA-RKKY}) Their ansatz for the self-energy
contains several free parameters which are 
fixed by fitting them to a
finite number of exactly calculated spectral moments. The temperature
dependence is contained implicitly in the magnetization which enters
the self-energy as an external parameter. In the limit of very low-$T$
we consider here, the average local moment is essentially unchanged
from its $T=0$ value, so the effects we uncover are basically absent
in the ansatz of Nolting et al.  In other words, besides studying
different Hamiltonians by very different means, our studies also focus
on very different regimes: very low $T$, in our work,
vs. medium and high $T$ in
Ref. \onlinecite{Nolting-zero-bandwith}. Needless to say, in the
absence of an exact solution it is likely that a collection of 
approximations valid in different regimes will be needed in order to
fully understand this problem.

The article is organized as follows: in Section II we
introduce our models and in Section III we derive the lowest-T
self-energy correction. Section IV presents our results and Section V
contains our conclusions.

\section{Models} \label{sec:Models}

We consider a single spin-${1\over 2}$ charge carrier which propagates
on a hypercubic lattice with periodic boundary conditions after $N_i$
sites in the direction $i=1,d$; the total number of sites is
$N=\prod_{i=1}^d N_i$. Our results are for $d=2$ and $d=3$. Of course,
long-range FM order at finite-T only exists in $d=3$. However, we also
consider anisotropic layered compounds, like the manganites, which
have 2D FM layers whose finite-T long-range order is stabilized by
weak inter-layer coupling, but where one can assume that at very
low-$T$ the intra-layer carrier dynamics determine its properties. In
principle, similar arguments can be employed to study $d=1$ chains
with FM order at finite-$T$ maintained by their immersion in 3D
lattices, but complications due to formation of magnetic domains
would still need to be  dealt with.

The carrier is an electron in an otherwise empty band
or a hole in an otherwise full band,  described by a tight
binding model with nearest neighbor (nn) hopping: 
\begin{align}
 \hat{T}=\sum_{\mathbf{k},\sigma} \epsilon(\mathbf{k})
 c^\dagger_{\mathbf{k},\sigma} c_{\mathbf{k},\sigma}, 
\end{align} 
with  $\epsilon(\mathbf{k})=-2t \sum_{i=1}^d \cos k_i$ for lattice
constant $a=1$.  
$c_{\mathbf{k},\sigma}^\dagger$ creates a carrier with momentum
$\mathbf{k}$ and spin $\sigma$.

The  local magnetic moments are described by either a
Heisenberg or an Ising interaction:
\begin{align}
 \hat{H}_{S}=-J \sum_{\langle i, j\rangle }\left ( \mathbf{S}_i\cdot \mathbf{S}_j-S^2 \right)
\end{align}
for Heisenberg exchange, while for Ising exchange:
\begin{align}
 \hat{H}_{I}=-J \sum_{\langle i, j\rangle }\left ( S^z_iS^z_j-S^2 \right),
\end{align} 
where $ \mathbf{S}_i$ is the spin-$S$ moment located at site
$\mathbf{R}_i$ and only nn exchange is included in both models. We represent local
moments  with a double arrow, {\em eg.}
$\Uparrow$, while the  carrier spin is represented by
a single arrow, {\em eg.} $\uparrow$. 

For both these models the undoped ground state is  $|{\rm
  FM}\rangle=\mid \Uparrow, \Uparrow, \dots\rangle$ and has zero
energy. The only excited states of
interest will be the single magnon states:
\begin{align}
 |\Phi(\mathbf{q})\rangle=\frac{S^-_{\mathbf{q}}}{\sqrt{2S}}|{\rm
   FM}\rangle=
 \sum_j\frac{\exponential{i\mathbf{q}\mathbf{R}_j}}{\sqrt{2SN}}S_j^-|{\rm
   FM}\rangle. 
\end{align}
Here $S^\pm_i=S^x_i\pm iS^y_i$ are the raising $(+)$ and lowering
$(-)$ operators. The key difference between the Heisenberg and Ising
interactions is the dispersion of the single magnon states. For the
Heisenberg model this is $\Omega_{\mathbf{q}}=4 J S\sum_{i=1}^d
\sin^2(q_i/2)$, whereas for the Ising model the magnons are
dispersionless, $\Omega_{\mathbf{q}}=\Omega=2dJS$.

The interaction between the carrier and the local moments is also a
Heisenberg exchange: 
\begin{align}
 \hat{H}_{\text{exc}}=J_0\sum_{j}\mathbf{s}_j \cdot \mathbf{S}_j,
\end{align}
where $\mathbf{s}_i=\sum_{\alpha,\beta} c_{i,\alpha}^\dagger
\frac{\mathbf{\sigma}_{\alpha,\beta}}{2} c_{i,\beta}$ is the carrier
spin operator and $\mathbf{\sigma}$ are the Pauli matrices. The coupling
$J_0$ can be either FM or AFM; we will consider both cases.

It is convenient to split
$\hat{H}_{\text{exc}}=\hat{H}_{\text{exc}}^z+\hat{H}_{\text{exc}}^{x,y}
$, where $\hat{H}_{\text{exc}}^z=J_0/2\sum_j \left (
  c_{j,\uparrow}^\dagger c_{j,\uparrow}-c_{j,\downarrow}^\dagger
  c_{j,\downarrow}  \right ) S^z_j$ and $\hat{H}_{\text{exc}}^{x,y}=J_0/2 \sum_j \left(
  c_{j,\uparrow}^\dagger c_{j, \downarrow}
  S^-_j+c_{j,\downarrow}^\dagger c_{j, \uparrow} S^+_j \right).$
The first term causes an energy shift $\pm
J_0 S/2$. The second term is responsible for spin-flip processes,
where the carrier flips  
its spin  by  absorbing or emitting a magnon. 

The total Hamiltonian is:
\begin{align}
 \hat{H}=\hat{T}+\hat{H}_{\text{S/I}}+\hat{H}_{\text{exc}}.
\end{align}
Due to translational invariance, the total momentum is
conserved. Furthermore, the $z-$component $S_{\text{tot}}^z$ of the
total spin (the sum of the carrier spin and lattice spins), is also
conserved. Therefore, eigenstates 
$\hat{H}|\psi^{(m)}_\alpha(\mathbf{k}) \rangle= E_\alpha^{(m)}(\mathbf{k})
|\psi^{(m)}_\alpha(\mathbf{k}) \rangle$ are indexed by the total momentum
of the system, $\mathbf{k}$, by the number $m$ of magnons when the
carrier has spin-up so that $S_{\text{tot}}^z=NS+\frac{1}{2}-m$, and
by $\alpha$ which comprises all the other quantum numbers.

\section{Formalism}
 \label{sec:method}

We want to calculate the low-T expression of  Zubarev's double-time
retarded propagator\cite{zubarev} for a spin-up carrier:
\begin{align}
 &G_{\uparrow}(\mathbf{k},\tau)=-\frac{i}{Z}\Theta(\tau)\tr
  [\exponential{-\beta \hat{H}} c_{\mathbf{k},\uparrow}(\tau)
    c_{\mathbf{k},\uparrow}^\dagger(0)], 
 \label{eq:GF-time}
\end{align}
but in a canonical (not grand-canonical) ensemble, assuming that
the carrier is injected in the otherwise undoped FM. As a result, the
trace is over the eigenstates of $\hat{H}_{\text{S/I}}$ (in the
absence of carriers, $\hat{H}\equiv
\hat{H}_{\text{S/I}}$). $\Theta(\tau)$ is the Heaviside function, $Z=
\tr [\exponential{-\beta \hat{H}_{\text{S/I}}}]$ is the partition
function for the undoped FM, and $
c_{\mathbf{k},\uparrow}(\tau)=\exponential{i \tau \hat{H}}
c_{\mathbf{k}, \uparrow} \exponential{-i \tau \hat{H}}$ is the carrier
annihilation operator in the Heisenberg picture. In the
frequency domain we have:
$$G_{\uparrow}(\mathbf{k},\omega)= \int_{-\infty}^{\infty} d \tau e^{i
  \omega \tau}G_{\uparrow}(\mathbf{k},\tau).
$$

At $T=0$, the trace reduces to a trivial expectation value over $|{\rm
  FM}\rangle$, and we find:\cite{Shastry}
$$
G^{(0)}_{\uparrow}(\mathbf{k}, \omega)=\langle {\rm FM}|
c_{\mathbf{k}, \uparrow} \hat{G}(\omega) c^\dagger_{\mathbf{k},
  \uparrow} |{\rm FM}\rangle = \frac{1}{\omega-E_{\uparrow}(\mathbf{k})+i \eta}.
$$
Here $\hat{G}(\omega)=[\omega-\hat{H}+i \eta]^{-1}$ is the resolvent
of $\hat{H}$ and $\eta$ is a small, positive number (we set
$\hbar=1$). Physically, 
$1/\eta$ sets the carrier lifetime.  
The  eigenenergy is
$E_{\uparrow}(\mathbf{k})=\epsilon(\mathbf{k})+J_0\frac{S}{2}$ for
both the
Heisenberg and Ising models. As discussed, this shows that at $T=0$ a
spin-up carrier 
propagates freely and acquires an energy shift from $\hat{H}_{\text{exc}}^z$.

At finite temperature, we expect to find:
\begin{align}
& G(\mathbf{k},
  \omega)=\frac{1}{\omega-E_{\uparrow}(\mathbf{k})-\Sigma(\mathbf{k},\omega)+i
    \eta} \nonumber \\ 
 &=G^{(0)}_{\uparrow}(\mathbf{k},
  \omega)+G^{(0)}_{\uparrow}(\mathbf{k}, \omega)
  \Sigma(\mathbf{k},\omega) G^{(0)}_{\uparrow}(\mathbf{k}, \omega) +
  \dots 
 \label{eq:self_energy_implicit} 
\end{align}
Strictly speaking, the energy shift $J_0\frac{S}{2}$ is part of the
self-energy, however it is 
convenient to separate it as we do here so that
$\Sigma(\mathbf{k},\omega)$ contains 
only the finite-$T$ terms.

Since we are interested in the lowest-$T$ contribution to
$\Sigma(\mathbf{k},\omega)$, we  consider only the
first two terms of Eq. (\ref{eq:GF-time}): 
\begin{align}
 G_{\uparrow}(\mathbf{k},\omega)=\frac{
   G_{\uparrow}^{(0)}(\mathbf{k},\omega) +  \sum_{\mathbf{q}}
   \exponential{-\beta \Omega_{\mathbf{q}}}
     G_{\uparrow}^{(1)}(\mathbf{k}, \mathbf{q}, \mathbf{q},\omega) +
     \dots }{1 + \sum_{\mathbf{q}}\exponential{-\beta
     \Omega_{\mathbf{q}}} + \dots },
 \label{eq:GF-freq}
\end{align}
where we define the new propagators $G_{\uparrow}^{(1)}(\mathbf{k},
\mathbf{q}, \mathbf{q}',\omega)= -i\int_{0}^{\infty} \mathrm{d}\tau \,
\exponential{i \omega \tau}  \langle \Phi(\mathbf{q}') |
c_{\mathbf{k},\uparrow}(\tau)
c_{\mathbf{k+q'-q},\uparrow}^\dagger|\Phi(\mathbf{q}) \rangle.$ Only
diagonal $\mathbf{q'}=\mathbf{q}$ terms contribute to the 
trace. After carrying out  the Fourier transform we find
$
 G_{\uparrow}^{(1)}(\mathbf{k}, \mathbf{q}, \mathbf{q}',\omega)=
 \langle \Phi(\mathbf{q'})|c_{\mathbf{k},\uparrow}
 \hat{G}(\omega+\Omega_{\mathbf{q'}})
 c_{\mathbf{k+q'-q},\uparrow}^\dagger |\Phi(\mathbf{q}) \rangle.$
Note that the argument of the resolvent  is
shifted by the magnon energy, meaning that the carrier's energy 
 is  measured with respect to that of the state in which the carrier is
 injected. Following calculations detailed in the Appendix, we find:

\begin{align}
& \sum_{\mathbf{q}}\exponential{-\beta \Omega_{\mathbf{q}}}
  G_{\uparrow}^{(1)}(\mathbf{k}, \mathbf{q}, \mathbf{q},\omega)=
  \sum_{\mathbf{q}} \exponential{-\beta \Omega_{\mathbf{q}}} \left \{
  G^{(0)}_{\uparrow}(\mathbf{k},\omega) \right. \nonumber \\
& \left. -\frac{J_0}{2N} \frac{
    [G^{(0)}_{\uparrow}(\mathbf{k},\omega)]^2 }{1+ J_0 SG^{(0)}_{\uparrow}(\mathbf{k+q},\omega+\Omega_{q})+\frac{J_0}{2} g(\mathbf{k},\mathbf{q},\omega)} \right \},
 \nonumber  \label{eq:GF-first-order-contrib}
\end{align}
where
$$
g(\mathbf{k},\mathbf{q},\omega)=\frac{1}{N} \sum_{\mathbf{Q}}
G^{(0)}_{\uparrow}(\mathbf{k+q-Q},\omega+\Omega_{\mathbf{q}}-\Omega_{\mathbf{Q}}) 
$$ is a known function. When this expression is used in Eq. (\ref{eq:GF-freq}), we
obtain
\begin{widetext}
\begin{align} G_{\uparrow}(\mathbf{k},\omega)&=\frac{
   G_{\uparrow}^{(0)}(\mathbf{k},\omega)(1 + \sum_{\mathbf{q}}\exponential{-\beta
     \Omega_{\mathbf{q}}}+ \dots)  +
   [G^{(0)}_{\uparrow}(\mathbf{k},\omega)]^2
   \Sigma(\mathbf{k},\omega)(1+\dots)+ \dots }{1+
    \sum_{\mathbf{q}}\exponential{-\beta 
     \Omega_{\mathbf{q}}}+ \dots}\nonumber \\
&=
  G_{\uparrow}^{(0)}(\mathbf{k},\omega) +[G^{(0)}_{\uparrow}(\mathbf{k},\omega)]^2
   \Sigma(\mathbf{k},\omega)+ \dots \nonumber,
\end{align} since the terms in the
brackets are the expansion of $Z$ (to the order considered here; higher order
contributions will come from including many-magnon processes) and cancel
with the denominator. This has the expected form of 
Eq. (\ref{eq:self_energy_implicit}), so we can identify the lowest-$T$
correction to the self-energy:
\begin{equation}
 \Sigma(\mathbf{k},\omega)=-\frac{J_0}{2N}  \sum_{q} \frac{
   \exponential{-\beta \Omega_{\mathbf{q}}} }{1+
   J_0SG^{(0)}_{\uparrow}(\mathbf{k+q},\omega+\Omega_{q})+\frac{J_0}{2}
   g(\mathbf{k},\mathbf{q},\omega)} +\dots. 
\label{se}
\end{equation}
\end{widetext}

It is important to mention that although we only considered states
with zero or one magnon in our  derivation, we will see some
higher-order effects in our results when using 
$G_{\uparrow}(\mathbf{k},\omega)=[\omega-E_{\uparrow}(\mathbf{k}) -
  \Sigma(\mathbf{k}, \omega) + i \eta]^{-1}$, {\em i.e.} when  the
self-energy is placed in the denominator. These are from states where
multiple magnons are 
present in the system but the carrier interacts only with  one
of them while the rest are ``inert'' spectators.  

Eq. (\ref{se}) is the main result of this work. The only difference
between Heisenberg and Ising backgrounds is the expression for the magnon energy
$\Omega_{\mathbf{q}}$. For the Ising case, this energy is independent of
momentum, resulting in a self-energy $\Sigma(\omega)$
independent of $\mathbf{k}$.

Before presenting results, let us consider what the spectral weight
$A_{\uparrow}(\mathbf{k}, \omega)= -\frac{1}{\pi}
\text{Im}G_{\uparrow}(\mathbf{k},\omega)$ should be expected to reveal. 
 The Lehmann representation of the
propagator in its expanded form is:
\begin{widetext}
\begin{align}
 &G_{\uparrow}(\mathbf{k},\omega)=\frac{1}{Z} \left [ \frac{1}{\omega
      - E_{\uparrow}(\mathbf{k})+i \eta}+ \sum_{\alpha, \mathbf{q}}
    \exponential{- \beta \Omega_{\mathbf{q}}} \frac{|\langle
      \Phi(\mathbf{q}) |c_{\mathbf{k},\uparrow}
      |\Psi_{\alpha}^{(1)}(\mathbf{k+q})\rangle|^2}{\omega
      +\Omega_{\mathbf{q}}-E^{(1)}_{\alpha}(\mathbf{k+q})+i \eta} + \dots
    \right ]. 
  \label{eq:Lehmann}
\end{align}
\end{widetext}
At $T=0$ only the first term contributes, giving a single
quasiparticle peak at $\omega = E_{\uparrow}(\mathbf{k})$. The second
term has poles 
at $\omega=E^{(1)}_{\alpha}(\mathbf{k+q})-\Omega_{\mathbf{q}}$. The $m=1$ 
subspace also corresponds to a spin-down carrier
injected in the FM at $T=0$, thus we can find the energies
$E^{(1)}_{\alpha}(\mathbf{k})$ from the spectral weight
$A_{\downarrow}^{(0)}(\mathbf{k}, \omega)= -\frac{1}{\pi}
\text{Im}G_{\downarrow}^{(0)}(\mathbf{k},\omega)$ where:
\begin{align}
G^{(0)}_{\downarrow}(\mathbf{k}, \omega)&=\langle {\rm FM}|
c_{\mathbf{k}, \downarrow} \hat{G}(\omega) c^\dagger_{\mathbf{k},
  \downarrow} |{\rm FM}\rangle \nonumber \\
& =\sum_{n}  \frac{|\langle {\rm FM} |c_{\mathbf{k},\downarrow}
  |\Psi_{\alpha}^{(1)}(\mathbf{k})\rangle|^2}{\omega -E^{(1)}_{\alpha}(\mathbf{k})+i
  \eta} \nonumber \\
& = \left \{ [ G^{(0)}_{\uparrow}(\mathbf{k},\omega)]^{-1} +\frac{J_0 S}{1+\frac{J_0}{2} g(\mathbf{k},0,\omega)}\right \}^{-1}
\end{align}
The last result is from Ref. \onlinecite{Shastry}. As already
mentioned and further detailed below, the spectrum
$E^{(1)}_{\alpha}(\mathbf{k})$ certainly contains an up-carrier+magnon
continuum spanning the energies $\{
E_{\uparrow}(\mathbf{k}-\mathbf{q'})+\Omega_{\mathbf{q'}}\}_{\mathbf{q'}}$;
in the right circumstances, a coherent spin-polaron state with the
magnon bound to the carrier may also appear, see below. Thus, for
$T\ne 0$, $A_{\uparrow}(\mathbf{k}, \omega)$ should have 
weight at all energies
$\{E_{\uparrow}(\mathbf{k+q-q'})+\Omega_{\mathbf{q'}}-
\Omega_{\mathbf{q}}\}_{\mathbf{q,q'}}$. In the Ising case the magnon
energies cancel out so weight should be expected at all energies $\{
E_{\uparrow}(\mathbf{q})\}_{ \mathbf{q}}$ in the spin-up carrier
spectrum, not just at $E_{\uparrow}(\mathbf{k})$. This automatically
implies that the $T=0$ infinitely lived quasiparticle of 
energy $E_{\uparrow}(\mathbf{k})$ acquires a finite lifetime at $T\ne
0$. This remains true for the Heisenberg case, with the added
complication that now, $\{E_{\uparrow}(\mathbf{k+q-q'})+\Omega_{\mathbf{q'}}-
\Omega_{\mathbf{q}}\}_{\mathbf{q,q'}}$ will generally span a wider
range of  energies than $\{E_{\uparrow}(\mathbf{q})\}_{
  \mathbf{q}}$. If a spin-polaron appears in the $m=1$ sector,
additional weight is expected at energies in its band minus the
magnon energy.
Higher order terms will contribute similarly (remember
that our solution for the propagator  does include partial
contributions from many-magnon states). To conclude,  at finite $T$
one can no longer assume that energies for which the spectral weight
$A_{\uparrow}(\mathbf{k}, \omega)$ 
is non-zero are necessarily in the spectrum of the
momentum-$\mathbf{k}$ Hilbert subspace. This makes the interpretation
of the spectral weight less straightforward than it is at $T=0$.

\section{Results}
 
\subsection{Review of $T=0$ results}
\label{sec:zero-temp}

Given the analysis presented above, it is useful to first quickly review the
dispersion $E_{\uparrow}(\mathbf{k})$ and, more importantly, the
spectrum $E^{(1)}_{\alpha}(\mathbf{k})$ for the $m=0$ and $m=1$
sectors, respectively. The latter is easiest to see by 
plotting the spectral weight $A_{\downarrow}^{(0)}(\mathbf{k},
\omega)$. The main focus will be to understand when a spin-polaron state
forms in the $m=1$ sector, but we will also verify the presence of the
continuum at the expected location. Since experimentally this is the
most relevant regime, we will assume that $|J_0|$ is the largest
energy scale and $J$ is the smallest one. While realistically one
expects $J \ll t$, we will set $J/t=0.5$ so that its role can be
discerned easily.

\begin{figure}[t]
 \includegraphics[angle=-90, width=0.45\textwidth]{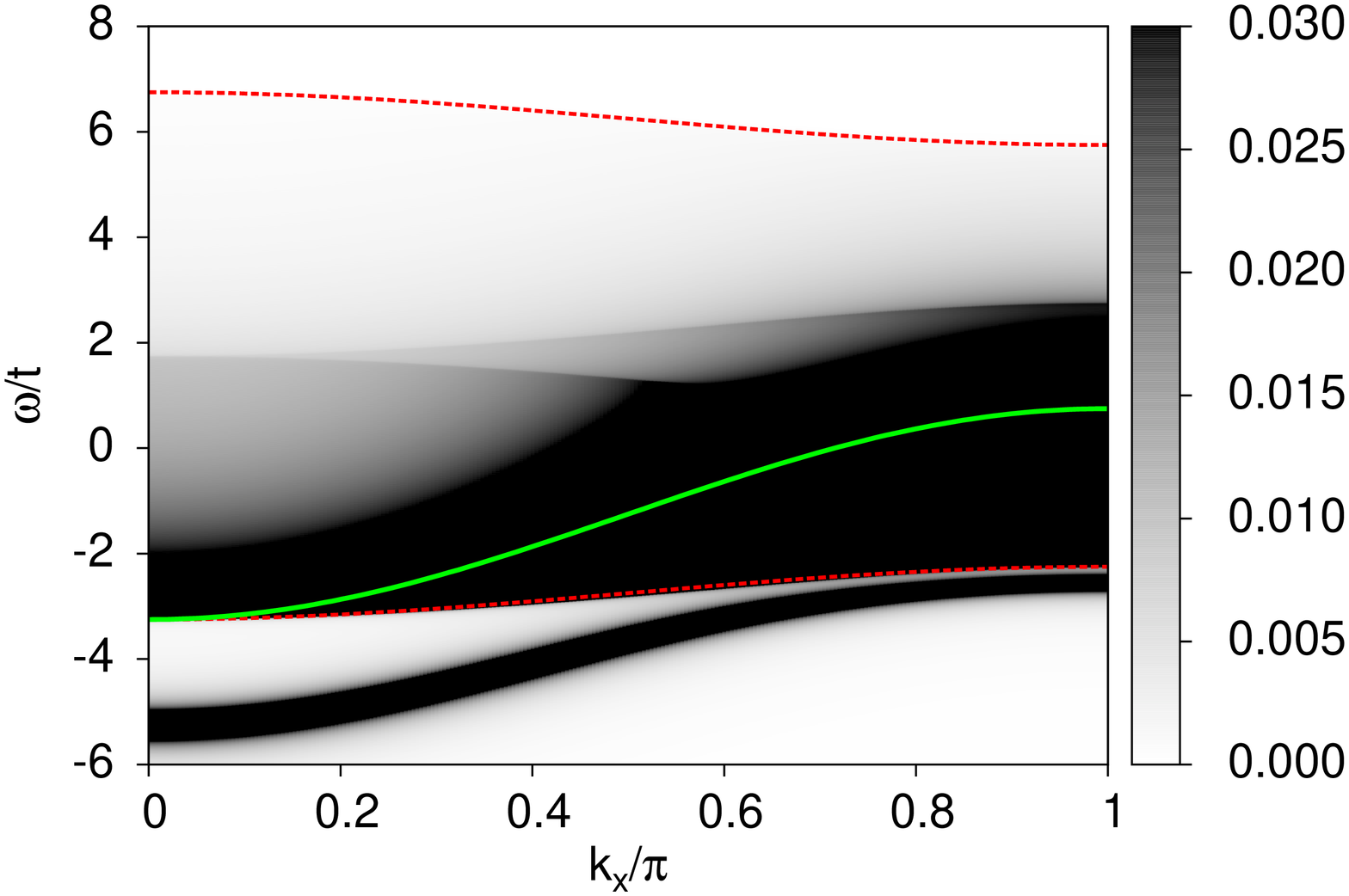}
 \includegraphics[angle=-90, width=0.45\textwidth]{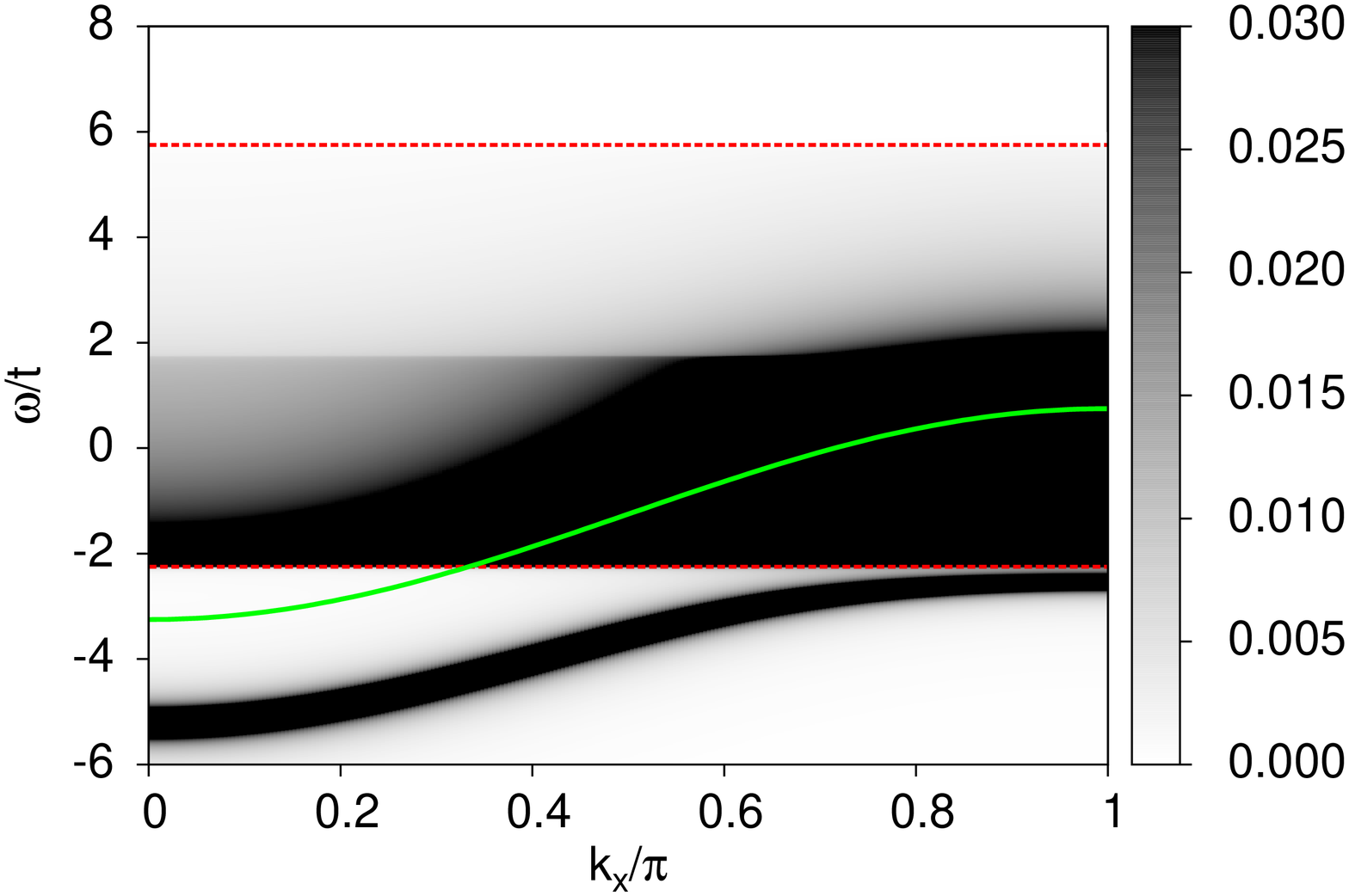}
 \caption{(color online) Energy $E_{\uparrow}(\mathbf{k})$ (thick full green line)
   and spectral weight  $A^{(0)}_{\downarrow}(\mathbf{k},\omega)$
   (contour map) vs. $k_x$ at $k_y=0$, 
   for the Heisenberg model (top) and the Ising model (bottom) in
   2D,  for AFM coupling $J_0/t=3$.  The dashed red lines mark
   the expected continuum boundaries in the $m=1$ subspace. 
 Other parameters are $J/t=0.5, S=0.5,  \eta/t=0.01$.}
 \label{fig:sp_map}
\end{figure}

Figure \ref{fig:sp_map} shows $E_{\uparrow}(k_x,k_y=0)$ (thick full
green line) and the spectral weight
$A^{(0)}_{\downarrow}(k_x,k_y=0,\omega)$ (contour map) for the 2D
Heisenberg and Ising models, for antiferromagnetic coupling
$J_0=3$. The spectrum of the $m=1$ sector consists of a discrete state
at low energies, the spin-polaron, and the up-carrier + magnon (c+m)
continuum at higher energies.  Because we will encounter a 
different spin-polaron later on, we will refer to this spin polaron as
``sp1''. To first order in perturbation theory its effective mass  is
a factor of $(2S+1)$ larger than the bare carrier mass and its 
energy is $-J_0(S+1)/2+ {\cal O}(t,J)$. \cite{Berciu-sp} Most of this
energy comes from $\hat{H}^z_{\text{exc}}$ and explains why for AFM
$J_0>0$ sp1 is the ground state -- states in the continuum have the
carrier with spin up and therefore cost $ \sim J_0S/2$ in exchange
energy. This also suggests that for FM coupling $J_0 <0$, the sp1
polaron should be located above the c+m continuum. This expectation is
confirmed below.

\begin{figure}[t]
 \includegraphics[angle=0, width=0.4\textwidth]{fig2a.eps}
 \includegraphics[angle=-90, width=0.4\textwidth]{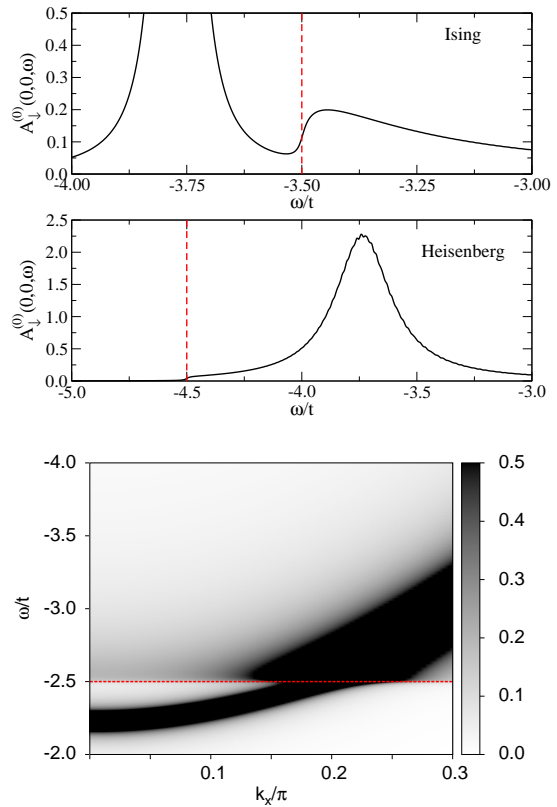}
 \caption{(color online) Top: 
   $A^{(0)}_{\downarrow}(\mathbf{k}=0,\omega)$ for FM
   $J_0/t=-2$ in 2D. The lower c+m continuum edge is marked with
   dashed red lines. The Ising model has a discrete state (sp2)
   below the continuum. Bottom: Spectral weight
   $A^{(0)}_{\downarrow}(\mathbf{k},\omega)$ for the Ising model in
   2D for $k_y=0, k_x<0.3\pi$. The dashed red line marks the lower c+m
   continuum edge. The sp2 state appears for small $\mathbf{k}$ and
then   merges into the continuum. Other parameters are $J/t=0.5,
S=0.5,  \eta/t=0.01$.} 
 \label{fig:peak_below_c+m}
\end{figure}

Comparing the two panels, we see that the sp1 dispersion is very
  similar for the two models. This is
 expected because this is a coherent state where the  
 magnon is locked into a singlet with the carrier, and this process is
 controlled by $J_0\gg J$. 
A difference appears in the shape of the c+m
continuum, however. As mentioned, this must span energies $\{
E_{\uparrow}(\mathbf{k}-\mathbf{q})+\Omega_{\mathbf{q}}\}_{\mathbf{q}}$
since it consists of up-carrier and magnon scattering states. 
The dashed red lines show the boundaries of this range, in agreement
with the data (this is more difficult to see for the upper edge, on
this scale, due to the reduced spectral weight at high energies).
Since Ising magnons are dispersionless the continuum boundaries do not
change with $\mathbf{k}$. In contrast, the continuum boundaries for
the Heisenberg model vary with $\mathbf{k}$, the continuum being wider
at the centre of the Brillouin zone than near its edges.

\begin{figure}[t]
 \includegraphics[angle=0, width=0.45\textwidth]{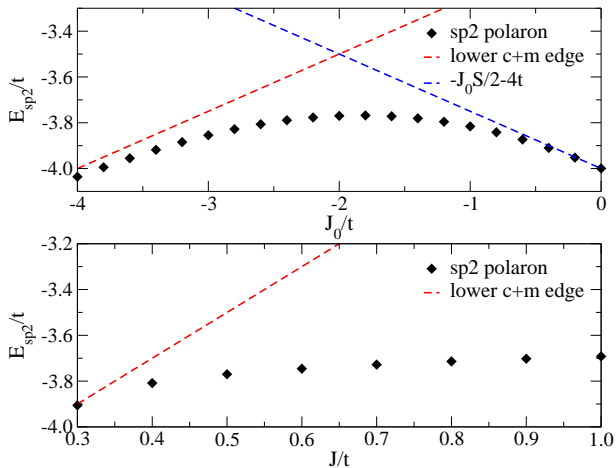}
 \caption{(color online) Ground-state energy of the Ising sp2 polaron
   as a function of 
   $J_0/t$ for $J/t=0.5$ (top) and as a function of  $J/t$ for
   $J_0/t=-2$ (bottom), for $S=0.5$.} 
 \label{fig:sp2_energy}
\end{figure}

This  difference  has consequences for a FM coupling $J_0
<0$. As mentioned, in this case the c+m continuum is expected to
be the low-energy feature in the $m=1$ spectrum, with the sp1 state
appearing above it. This is indeed the case for the Heisenberg model,
however in the Ising model, for a sufficiently large $J$, a second
discrete state emerges below the c+m continuum.  We will refer to this
state as ``sp2''  to distinguish it from sp1.  
The top panel of Fig. \ref{fig:peak_below_c+m} shows its
presence (absence) for the Ising (Heisenberg) model at
$\mathbf{k}=0$. The bottom panel shows that even for the Ising model,
the sp2 only exists for small $\mathbf{k}$, at
least for these parameters. 

The origin of the sp2 state is suggested by the findings of Henning
et al. who showed that for $J=0$,  polaron-like states
exist inside the c+m continuum.\cite{Nolting} We believe that the addition of
$\hat{H}_{\text{I}}$ pushes one of them below the
continuum. This is possible because for an Ising coupling, the lower
continuum edge moves up by  $\Omega=2dJS$, whereas the polaron-like
states experience a smaller energy shift since they include a
component with the carrier having spin-down. For the Heisenberg model,
on the other hand, inclusion of $\hat{H}_{\text{S}}$ does not
change the location of the lower continuum edge at $\mathbf{k}=0$
since $\Omega_{\mathbf{q}=0}=0$, so the polaron-like state remains a
resonance inside the continuum.

\begin{figure}[t] 
\includegraphics[angle=-90, width=0.45\textwidth]{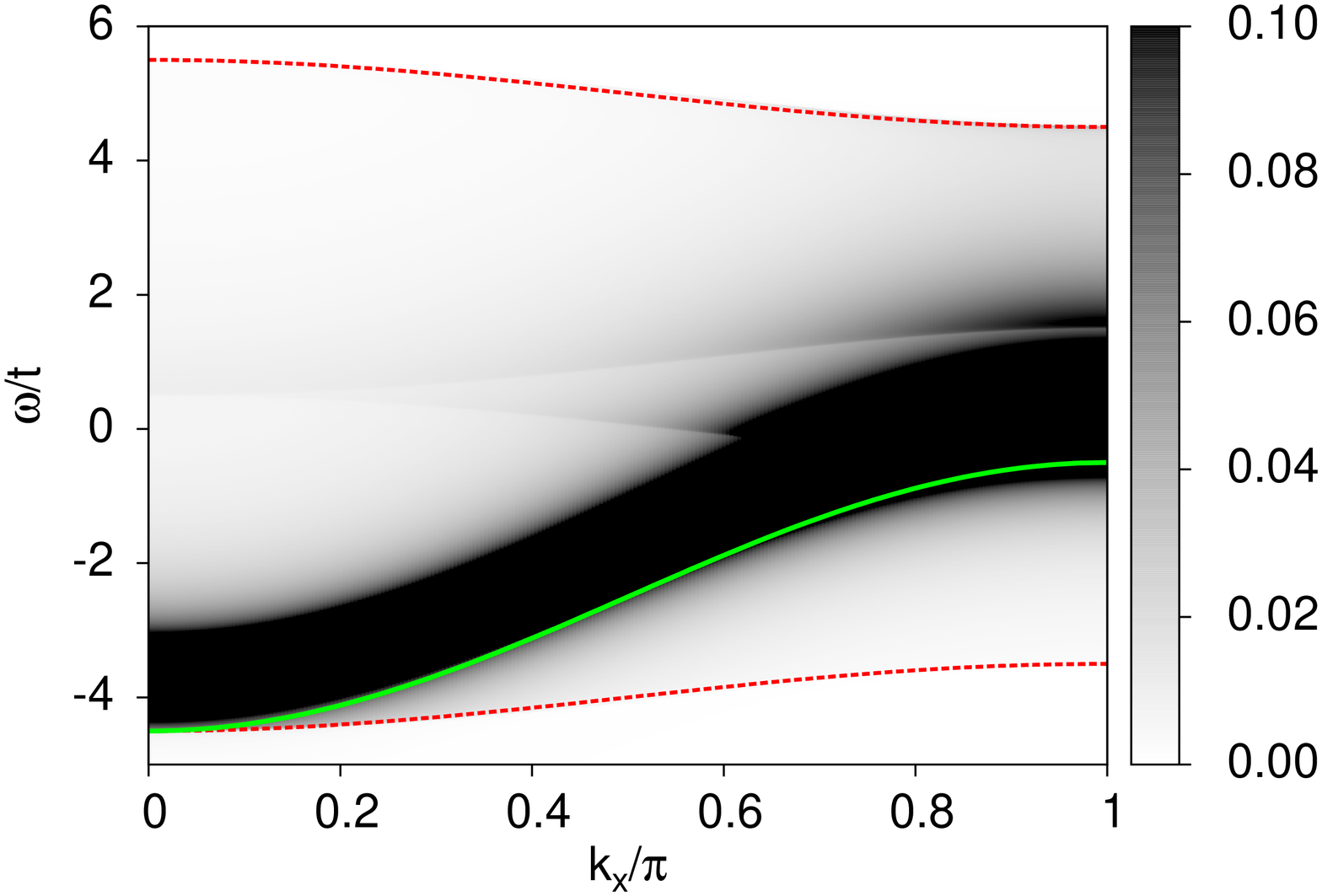}
\includegraphics[angle=-90, width=0.45\textwidth]{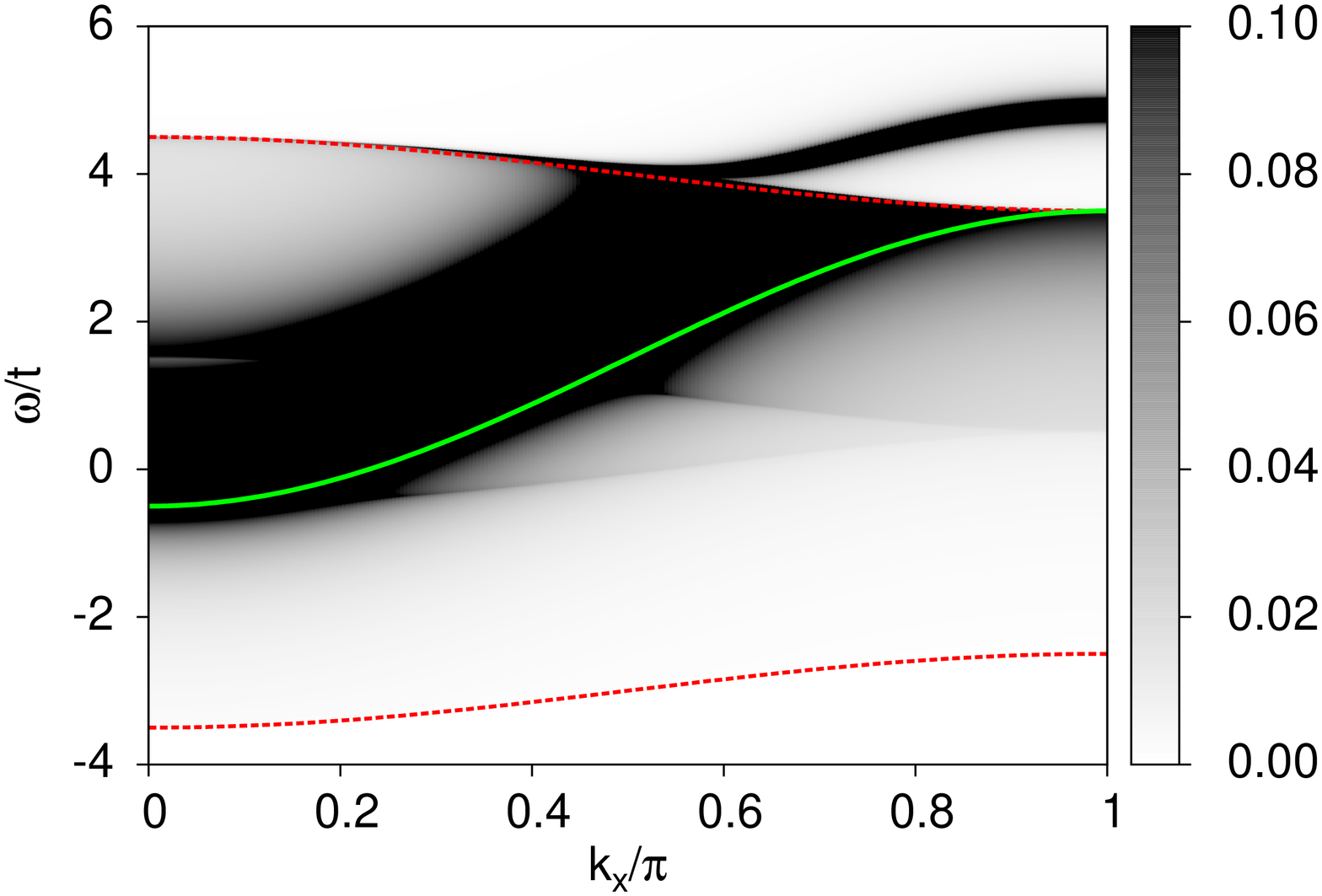}
 \caption{ (color online) Spectral weight
   $A^{(0)}_{\downarrow}(\mathbf{k},\omega)$ vs $k_x$ for 
   the 2D Heisenberg model at $k_y=0$ (top) and
   $k_y=\pi$ (bottom) and FM $J_0/t=-2$. Sp1 appears above
   the continuum only near the Brillouin zone edge. No sp2 peak is seen
   below the continuum. The dashed red
   lines mark the c+m 
   continuum boundaries and the green line marks
   $E_{\uparrow}(\mathbf{k})$. Other parameters are $J/t=0.5, S=0.5,
   \eta/t=0.01$.}  
 \label{fig:sp_map_jo_neg_2}
\end{figure}

The ground-state energy of the sp2 polaron is explored in
Fig. \ref{fig:sp2_energy}. The top panel shows its dependence on
$J_0/t$. The sp2 state  has weight on both the down-carrier 
and on the up-carrier+magnon components. For $J_0=0$ the weight of the
latter component must vanish since no spin-flips are possible and the
sp2 state is the same as a free down-carrier, whose energy
$-J_0S/2-4t$ is also indicated (dashed blue line). These results
suggest that as $|J_0|/t$ increases, the  sp2 state shifts weight from
the down-carrier component to the up-carrier+magnon component until it
essentially becomes a continuum-like state. 

The bottom panel in Fig. \ref{fig:sp2_energy} shows the sp2  ground-state
energy vs. $J/t$ for fixed $J_0/t=-2$. This value of
$J_0/t$ was chosen because  here the polaronic
character of sp2 is especially strong since if we neglect
$H_{\text{exc}}^{x,y}$, the energy of the down-carrier component is
equal to that of the up-carrier+magnon component. The distance between
sp2 and the continuum  increases with $J/t$, as expected
from our previous discussion.

While we have only seen the sp2 polaron for the Ising model, we cannot
rule out the possibility that for a very narrow range of momenta and
carefully chosen parameters, an sp2 state might also appear in the
Heisenberg model. Another important point is that the sp1 state is not
guaranteed to exist for all $\mathbf{k}$, either. In
Fig. \ref{fig:sp_map_jo_neg_2} we show
$A^{(0)}_{\downarrow}(\mathbf{k},\omega)$ for the 2D Heisenberg
model. No sp2 state appears below the continuum, and 
sp1 separates above the continuum only near the Brillouin zone
edge. This is not a surprise given the rather small value of $|J_0|$,
since it controls the separation between sp1 and
the continuum. For sufficiently large $|J_0|$, the sp1 polaron splits
off the continuum in the entire Brillouin zone.\cite{Shastry}

To summarize, the spectrum in the $m=1$ (one-magnon) subspace contains
the expected c+m continuum. For AFM $J_0$ the low-energy feature is
the sp1 polaron for both the Heisenberg and the Ising models. For FM
$J_0$, sp1 becomes the high energy feature and may only appear in a
small region of the Brillouin zone if $|J_0|$ is small. For the Ising
model and FM $J_0$, an sp2 polaron is also found to appear below the
c+m continuum, in a central region of the Brillouin zone that
increases with increasing $J$. For the Heisenberg model and FM $J_0$
we cannot entirely rule out the existence of sp2, although we provided
arguments which suggest that this is unlikely.

\begin{figure}[t]
 \includegraphics[angle=0, width=0.45\textwidth]{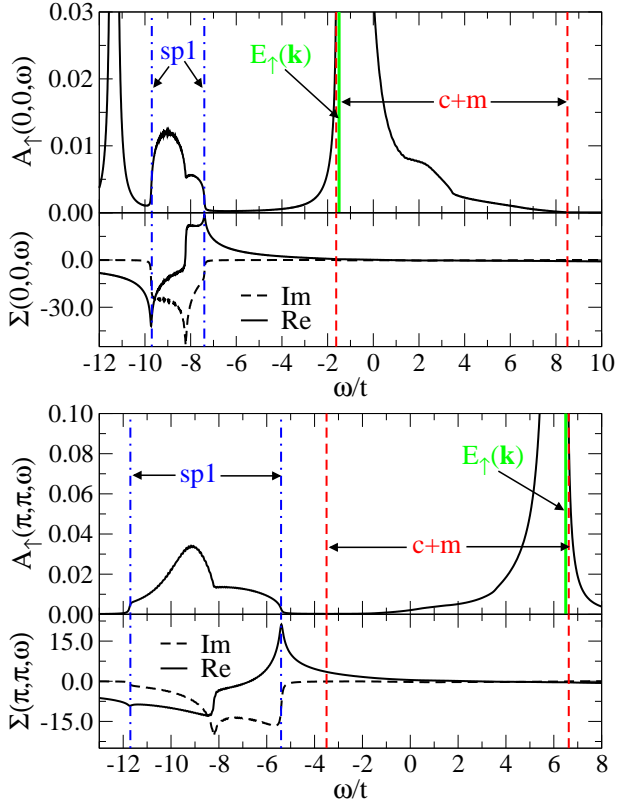}
 \caption{(color online) Spectral weight
   $A_{\uparrow}(\mathbf{k},\omega)$\ and the real (solid line) and imaginary 
   (dashed line) part of the self-energy $\Sigma(\mathbf{k},\omega)$ 
   for the 2D Heisenberg model with AFM $J_0/t=10$ and $\beta t=1$, at
   $\mathbf{k}=(0,0)$ (top) and $\mathbf{k}=(\pi,\pi)$ (bottom). The expected
   sp1 continuum 
   boundaries are marked with dash-dotted blue lines and the expected
   c+m continuum boundaries with dashed red lines. The
   $E_{\uparrow}(\mathbf{k})$  energy of the
   $T=0$ $\delta$-peak is marked with a thick green line. Other
   parameters are $J/t=0.5, S=0.5, \eta=0.02$ (top) and $\eta=0.05$ (bottom).}
 \label{fig:spectrum_jo_afm_H}
\end{figure}

We focused here more on the sp2 polaron because, to our knowledge, this
solution had not been discussed before, while the sp1
state has been analyzed in great
detail.\cite{Shastry,Berciu-sp,Nolting, Nakano}
We also note that while we 
presented only  (computationally less
costly to generate) 2D results, we find qualitatively similar results in 3D. This
will become clear from our finite-T results shown below.

\subsection{Low-T results} \label{sec:low-T}

We now present and analyze low-T results for the spectral weight of
the spin-up carrier. Since the calculation of
$G_{\uparrow}(\mathbf{k},\omega)$ becomes numerically very expensive
in 3D, most of our analysis is in 2D. However, we will also show a
selection of 3D spectra which prove that the 3D results are
qualitatively similar to the 2D results.

\begin{figure}[t]
 \includegraphics[angle=0, width=0.45\textwidth]{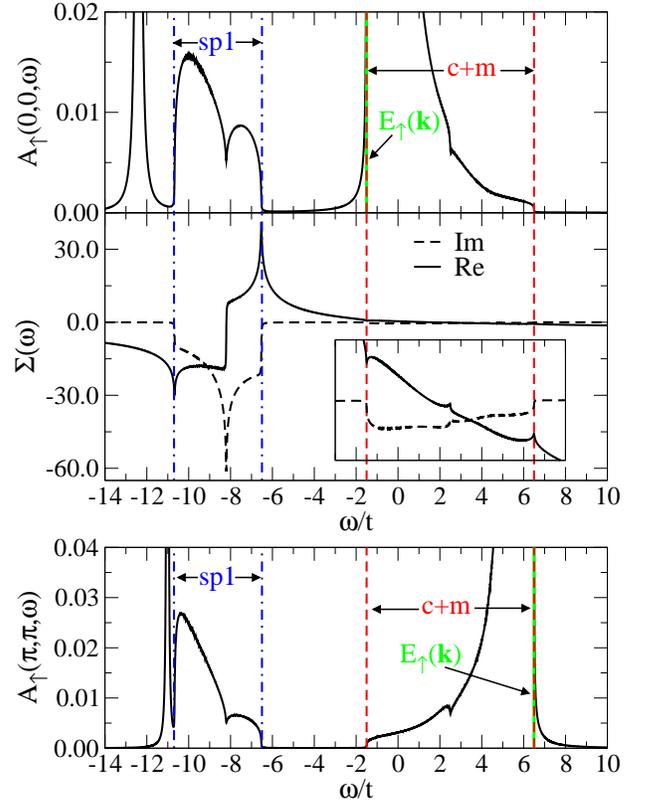}
 \caption{(color online) Same as Fig. \ref{fig:spectrum_jo_afm_H} but
 for the Ising model. All parameters are the same except $\beta t=0.5$
 and $\eta=0.01$ in both panels. Note that for the Ising model $\Sigma(\omega)
 $\ is independent of $\mathbf{k}$. The inset shows a zoom on $\Sigma(\omega)$\
 at high energies.}
 \label{fig:spectrum_jo_afm_I}
\end{figure}

The spectral weight $A_{\uparrow}(\mathbf{k},\omega)=-\frac{1}{\pi}
\text{Im}G_{\uparrow}(\mathbf{k},\omega)$ and the self-energy $\Sigma(\mathbf{k},\omega)$
are shown for the Heisenberg
and Ising models with AFM coupling $J_0/t=10$ in
Figs. \ref{fig:spectrum_jo_afm_H} and \ref{fig:spectrum_jo_afm_I},
respectively. In both cases the top panel is for $\mathbf{k}=(0,0)$
and the bottom one is for $\mathbf{k}=(\pi,\pi)$. {However, for the Ising model
the self-energy is independent of $\mathbf{k}$\ and therefore in Fig. \ref{fig:spectrum_jo_afm_I}
it is only shown beneath the $k=(0,0)$\ spectral weight.
The value of $J_0/t$
was chosen so large in order to ensure that the different features in
the spectrum are well separated, to simplify the analysis. Results for
smaller values of $J_0$ will be shown below.

 $A_{\uparrow}(\mathbf{k},\omega)$, which at $T=0$ is the peak
$\delta(\omega-E_{\uparrow}(\mathbf{k})) $ (indicated by the thick green
line), broadens into a continuum at finite-$T$. As
discussed at the end of Section III, this continuum has its origin in
the c+m continuum 
of the $m=1$ sector, thus we continue to call it the ``c+m''
continuum, and should span  $\{E_{\uparrow}(\mathbf{k+q-q'})+\Omega_{\mathbf{q'}}-
\Omega_{\mathbf{q}}\}_{\mathbf{q,q'}}$. The red dashed lines show the
boundaries of this energy range, in excellent agreement with the
broadening observed in $A_{\uparrow}(\mathbf{k},\omega)$. We 
note that most of the spectral weight is still located near
$E_{\uparrow}(\mathbf{k})$. 

This broadening confirms that at finite-T the quasiparticle acquires a
finite lifetime (the peak at $E_{\uparrow}(\mathbf{k})$ is now a
resonance inside a broad continuum, not a discrete state). Clearly,
this is due to processes where the spin-up carrier
absorbs a thermal magnon and then re-emits it with a different
momentum, thus scattering out of its original state.

The finite lifetime of the carrier in the c+m
continuum is also evident in the self-energy. The inset in Fig. 
\ref{fig:spectrum_jo_afm_I} shows that for energies within the 
c+m continuum the imaginary part of the self-energy is finite. The
same is true for the Heisenberg model (not shown).

While the broadening of the T=0 $\delta$-peak may be thought of as quite trivial,
Figs. \ref{fig:spectrum_jo_afm_H} and \ref{fig:spectrum_jo_afm_I} show
that it is not the only effect of the finite-T: spectral weight is also
transfered to a new continuum located below the c+m continuum. We
attribute this continuum to the sp1 state. Indeed, if we denote by
$E_{\text{sp1}}(\mathbf{k})$ the energy of the sp1 polaron, we find that this
continuum spans
$\{E_{\text{sp1}}(\mathbf{k+q})-\Omega_{\mathbf{q}}\}_{\mathbf{q}}$
(the boundaries of this range are marked by the dashed-dotted blue
lines). Its presence agrees with the Lehmann representation and
reveals this spectral weight transfer to be due to processes where the
spin-up carrier binds a thermal magnon and turns into an sp1 polaron.

The sp1 continuum is also where both the real and imaginary
part of $\Sigma(\mathbf{k},\omega)$\ take their largest values. Consequently 
the lifetime of these states is roughly two orders of magnitude smaller than 
that of the states within the c+m continuum. This is not surprising as the c+m
continuum stems from a $\delta$-peak with an infinite lifetime at T=0, 
whereas the sp1 continuum vanishes at T=0.

There is furthermore a qualitative difference between the real-part 
of $\Sigma(\mathbf{k},\omega)$\ in the sp1 continuum and in the c+m continuum.
For the latter the real part falls off relatively smoothly ({\em cf.} inset in Fig.
 \ref{fig:spectrum_jo_afm_I}), whereas for the sp1 continuum it is highly
singular and almost discontinuous.

Note that there are no major differences between the
Heisenberg and Ising models, except for the fact that the boundaries of
these continua are momentum dependent for the former and momentum
independent for the latter, due to their different magnon dispersions.

\begin{figure}[t]
 \includegraphics[angle=0, width=0.49\textwidth]{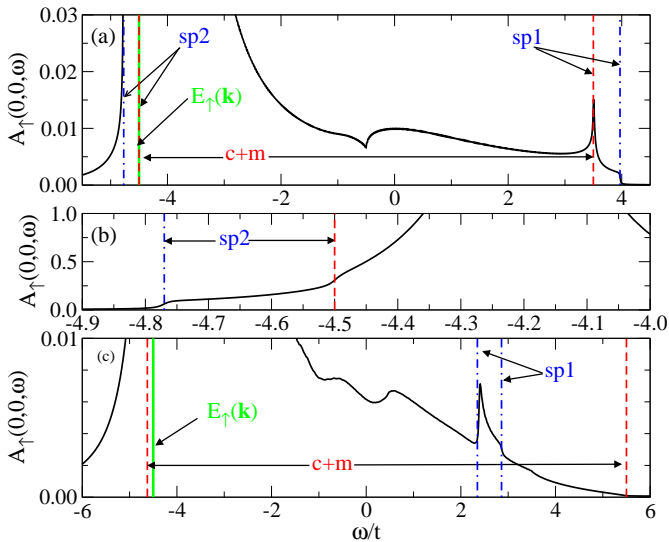}
 \caption{(color online) Spectral weight
   $A_{\uparrow}(0,0,\omega)$ for the 2D Ising model (panels
   (a) and (b)) and 2D Heisenberg model (panel (c)) for ferromagnetic
   $J_0/t=-2$ at $\beta t =0.5, \eta/t=0.01$ (Ising) and $\beta t=1,
   \eta/t=0.02$ (Heisenberg). The expected location of various features
   are also indicated (see text for more details). Other parameters
   are $J/t=0.5, S=0.5$.}
 \label{fig:spectrum_jo_fm}
\end{figure}

Figures \ref{fig:spectrum_jo_afm_H} and \ref{fig:spectrum_jo_afm_I} also
show a very puzzling discrete state at low energies. Before we turn
our attention to the analysis of this peak, we quickly discuss the
case with FM coupling $J_0<0$. Ising and Heisenberg results are
depicted in Fig. \ref{fig:spectrum_jo_fm} for $J_0/t=-2$ and
$J/t=0.5$. From the discussion of the $T=0$ spectrum in the $m=1$
Hilbert space, we know that for these parameters the Ising model has
an sp2 state below its c+m continuum and therefore expect to find its
signature in the finite-T spectrum, as well. This is indeed the case,
as seen more clearly in panel (b) which expands the low-energy part of
the Ising spectrum shown in (a), revealing weight at energies 
spanning
$\{E_{\text{sp2}}(\mathbf{k+q})-\Omega_{\mathbf{q}}\}_{\mathbf{q}}$
(its lower boundary is marked by dashed-dotted blue lines). Note that
since the sp2 state merges with the c+m continuum (boundaries marked
by red dashed lines), their corresponding continua also merge, but
panel (b) reveals a clear discontinuity where they overlap. The
high-energy sp1 continuum is also clearly observed in panel (a), again
merged with the c+m continuum since the sp1 state is not  fully
separated at such a small $|J_0|$, either.

The Heisenberg model (panel (c)) only shows the c+m and sp1 continua,
since there is no sp2 polaron here. Again, agreement with the
expected boundaries is excellent (the weight seen
below the c+m lower edge is due to the finite $\eta$ and the fact
that we zoomed in close to the axis to make it easier to see the sp1
continuum). 

It is worth noting that since for small $|J_0|$ the various
features merge, it would be easy to misinterpret the thermal
broadening as being all of c+m origin, {\em i.e.} to entirely miss the
role played by the spin-polaron solutions in the $m=1$ subspace. This
is also illustrated in Fig. \ref{fig:peak}, where we return to an AFM
$J_0$ coupling and show how the $\mathbf{k}=0$ spectra change as $J_0$
is decreased. All features discussed previously can be easily
identified for large $J_0$ but merge into one another as $J_0$
decreases, so that by the time $J_0/t=3$ there is only one very broad
feature, albeit with a non-trivial structure, left in the spectrum
(apart from  the low-energy discrete peak, which we will discuss
later). If
one assumed that this is all of c+m origin, {\em i.e.}  scattering of
the carrier on individual thermal magnons, one would infer very wrong
values of the parameters from the boundaries' locations.

\begin{figure}[t]
 \includegraphics[angle=0,  width=0.49\textwidth]{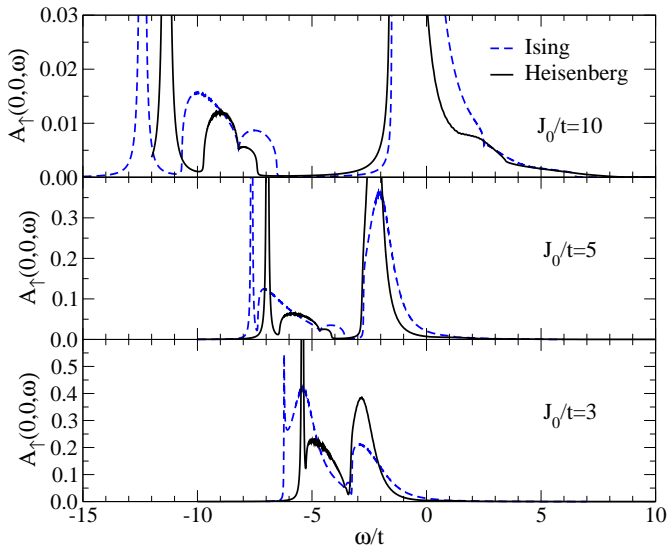} 
 \caption{(color online) Spectral weight
   $A_{\uparrow}(0,0,\omega)$ for the 2D Ising (dashed lines)
   and Heisenberg (full lines) models for $J_0/t=10, 5, 3$ in the top,
   middle and bottom panels, respectively. Other parameters are
   $J/t=0.5, S=0.5$ and $\beta t= 0.5, \eta/t=0.01$ (Ising), and
   $\beta t =1, \eta/t=0.02$ (Heisenberg). The oscillations visible
   especially in the sp1 continuum are due to finite-size effects (we
   used $N=100^2$ and $N=500^2$ for Heisenberg and Ising models, respectively).}
 \label{fig:peak}
\end{figure}

\begin{figure}[t]
 \includegraphics[width=0.4\textwidth]{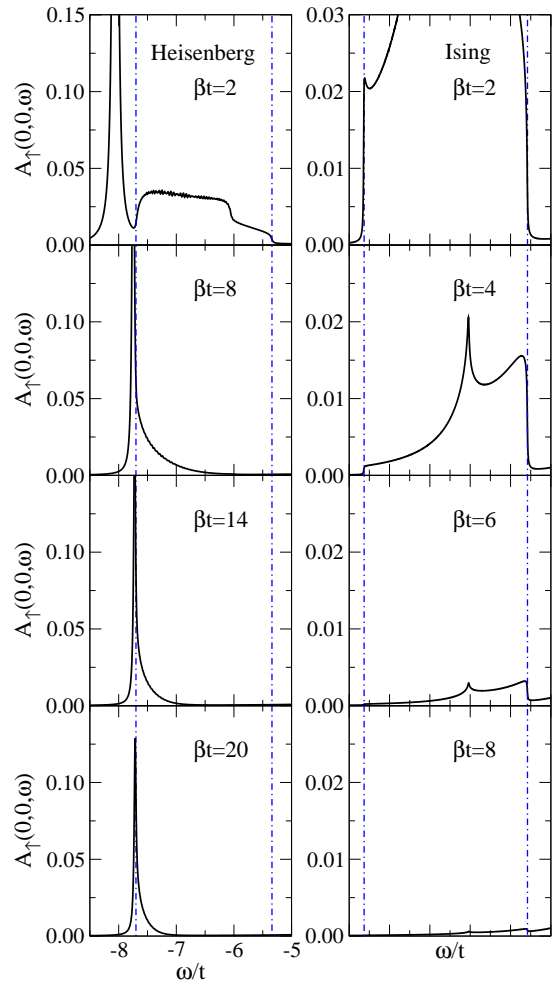}
 \caption{(color online) Spectral weight
   $A_{\uparrow}(\mathbf{k}=0,\omega)$ for the 
   2D Heisenberg (left) and Ising (right) models with AFM
   $J_0/t=7$, at different temperatures. Only the sp1 continuum is shown.
   Its edges are indicated with dot-dashed blue lines Other parameters are
   $J/t=0.5,  S=0.5$ and $\eta/t=0.01$ and
   $0.02$ for Ising and Heisenberg, respectively.}
 \label{fig:temp_dependence1}
\end{figure}

The results shown so far are for  large 
temperatures $k_B T \sim t= 2J$ (for our parameters), where higher
order corrections should certainly become quantitatively important. On the
other hand, from the Lehmann decomposition we expect that the location
of the various features does not depend on temperature; only how much
spectral weight they carry can change with $T$. For a more thorough
analysis we return to the case of AFM $J_0$, using a rather large
value so that the various features are well separated, and plot in
Fig. \ref{fig:temp_dependence1} the spectral weight in the sp1
continuum for several different temperatures, for both the Ising and
Heisenberg models. This confirms that, indeed, the weight in this
continuum decreases fast as $T\rightarrow 0$ while its location is not
affected (the location of the low-energy peak shifts with $T$, but as
we argue below, we do not believe that this is a physical feature). 

To quantify the spectral weight transferred, we calculate
$\int_{\text{c+m}} \mathrm{d}  
\omega   A_{\uparrow}(\mathbf{k},\omega)$, {\em i.e.} how much  is in the
c+m continuum. Since at $T=0$ all the
weight is in the
$\delta$-peak at $E_{\uparrow}(\mathbf{k})$ located inside the c+m
continuum, this value starts at 1 and decreases with increasing $T$,
as weight is transferred into the sp1 continuum; one can easily check that
the spectral weight obeys the sum rule $\int_{-\infty}^{\infty} \mathrm{d}
\omega   A_{\uparrow}(\mathbf{k},\omega)=1$.

\begin{figure}[t]
 \includegraphics[width=0.45\textwidth]{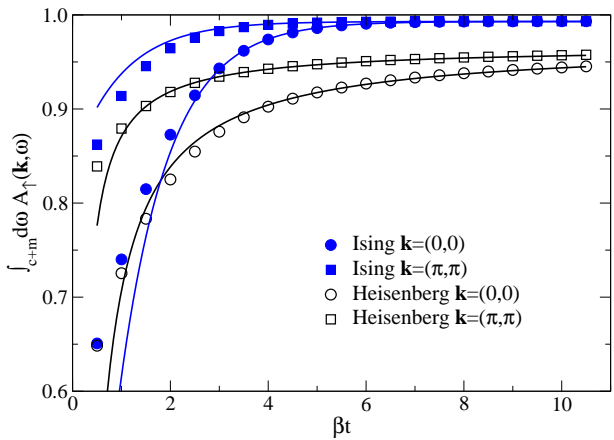}
 \caption{(color online) 
   Integrated spectral weight in the c+m continuum as a function of
   $\beta$. Lines are fits described in the text. Parameters are:
   $J_0/t=10, J/t=0.5, S=1/2, \eta/t=0.01$ (Ising), $\eta/t=0.05$
   (Heisenberg).} 
 \label{fig:temp_dependence}
\end{figure}

The results are shown in Fig. \ref{fig:temp_dependence} for both
models, both at the center and at the corner of the Brillouin
zone. Note that because of the finite value of $\eta$, some spectral
weight ``leaks'' outside the continuum's boundaries. This problem is
more severe at lower $T$ because $E_{\uparrow}(\mathbf{k})$ is located
very close to an edge of the continuum; this explains why the value
saturates below 1 as $\beta \rightarrow \infty$. This explanation is
also consistent with the observation that the amount of ``missing
weight'' as $T\rightarrow 0$ is of order $\eta$.

Two features are immediately apparent. First,  there is a substantial
difference in the amount of spectral weight transferred out of the c+m
continuum  at  $\mathbf{k}=(0,0)$
vs. $\mathbf{k}=(\pi,\pi)$. This is expected for the
Heisenberg model  where the location of all features changes with
$\mathbf{k}$, but may come as a surprise for the Ising model where their
location is independent of $\mathbf{k}$. However, for both models
$E_{\uparrow}(\mathbf{k})$, where most of the weight is found, moves
from the lower edge of the c+m continuum when 
$\mathbf{k}=0$, to the upper edge for $\mathbf{k}=(\pi,\pi)$. As a
result, it is reasonable that weight is transferred into the low-energy
sp1 continuum more efficiently  at $\mathbf{k}=(0,0)$ than at
$\mathbf{k}=(\pi,\pi)$, since in the former case the ``effective''
 energy difference between the two features is
smaller. 

The second observation is that spectral weight is transferred into the
sp1 continuum more efficiently in the Heisenberg model than in the
Ising model. This difference is also clearly visible
in Fig. \ref{fig:temp_dependence1}, where the weight in the sp1
continuum of the Heisenberg model is still respectable at $\beta t
=20$, while for the Ising model this weight is already negligible at
$\beta t =8$.

An explanation for this difference comes from assuming that
the weight in the sp1 continuum is proportional to the average number
of thermal magnons, since no sp1 polaron can appear in
their absence. Because the Ising magnon spectrum is gapped, at low-$T$
this number is proportional to the Boltzmann factor
$\exponential{-\beta \Omega}$. This suggests an integrated weight in
the c+m spectrum of  $a - b 
\exponential{-\beta 4JS}$, where $a=1- \cal{O}(\eta)$ is the limiting
value as $T\rightarrow 0$. We fitted the data points for $\beta t >5 $
with this form and found a very good fit (solid lines), which
moreover works well for a larger range of $\beta$ values than used in
the fit.

Magnons of the Heisenberg model are gapless so their number increases
much faster with $T$. A simple estimate for a 2D unbounded parabolic
dispersion suggests $\langle n \rangle \sim k_BT$.\cite{note} The
lines shown for the Heisenberg model in Fig. \ref{fig:temp_dependence}
are fits to $a-b/\beta$ for the data points with $\beta t > 5$. The
fit is again reasonable over a wider range, and much superior to other
simple functional forms we tried, such as $a-b/\beta^n$, $n >1$ or $a - b
\exponential{-\beta cJ}$ (the former assuming that we misindentified
the power law, the second to see if Ising-like fits might be more
appropriate). Of course, one can find excellent fits for all data
using more complicated functions with additional parameters,
but they are much harder to justify physically than
our simple hypothesis resulting in an  effectively one parameter fit.

\begin{figure}[t]
 \includegraphics[width=0.45\textwidth]{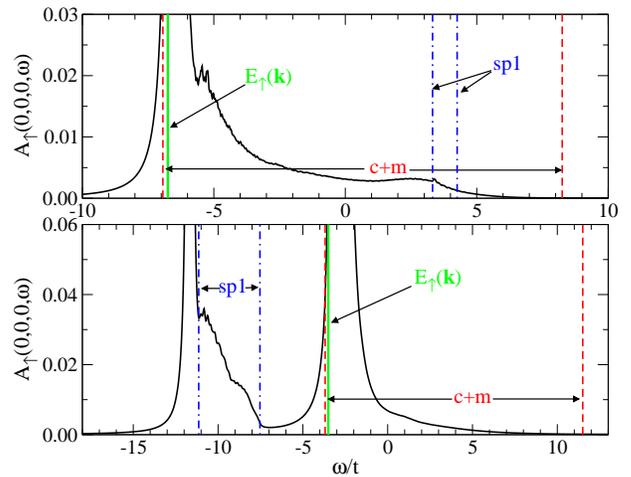}
  \caption{(color online) Spectral weight
    $A_{\uparrow}(\mathbf{k}=0,\omega)$ for 
    the 3D Heisenberg  model at $\beta t =1$ for  FM $J_0/t=-3$ (top) and AFM
    $J_0/t=10$ (bottom) couplings. The edges of the c+m 
   continuum (dashed red lines) and sp1/sp2 continuum (dot-dashed blue
   lines) are indicated, as is $E_{\uparrow}(0)$ (thick green
   line). Other parameters are
    $J/t=0.5,  S=0.5, \eta=0.1$. } 
 \label{fig:3dH}
\end{figure}

Let us now discuss the discrete peak appearing below the sp1 continuum
for both models, for AFM $J_0$. After carefully investigating many of its properties,
such as how its energy and the region in the Brillouin zone where it
exists depend on various parameters including $T$,\cite{extra}
we believe that this is an unphysical artefact of our
approximation. Arguments for this are: (i) the temperature dependence
of its location, clearly visible in Fig. \ref{fig:temp_dependence1}
(note that for the Ising model, the peak only separates below the sp1
continuum at higher $T$. At $\beta t =2$ one just starts to see weight
piling up near the lower edge, in preparation for this). According to
the Lehmann decomposition, the ranges where finite spectral weight is
seen cannot vary with $T$; (ii) the fact that the problem is worse at
higher-$T$, where we know that higher order corrections ought to be
included in the self-energy; these could easily remove an unphysical
pole; (iii) the fact that this is a discrete peak, not a resonance
inside a continuum (this can be easily verified by checking that its
lifetime is set by $\eta$). According to the Lehmann
decomposition, discrete peaks cannot appear in the $T\ne 0$ spectral
weight. Even if the carrier binds all thermal magnons in a coherent
quasiparticle, the finite-$T$ spectral weight would reveal only a 
continuum associated with it, as is the case for the sp1 and sp2
polarons. To summarize, we believe that this discrete peak is an
artefact and that in reality, its weight is
part of the sp1 continuum from which it came.

Ideally, these arguments would be strengthened by a calculation of the
next correction to the self-energy, to check its effects. We found the
exact calculation of the two-magnon term to be daunting even for the
Ising model. The difficulty is not so much in evaluating different
terms, but in tracing over all possible contributions -- so far we did
not find a sufficiently efficient way to do this. One can use
approximations to speed things up, 
but  that defeats the purpose since it would not be clear if the end
results are intrinsic or artefacts, as well. 
Given this, we
cannot entirely rule out that the discrete peak is a (precursor
pointing to a) real feature, but we believe that to be very unlikely.

\begin{figure}[t]
 \includegraphics[width=0.44\textwidth]{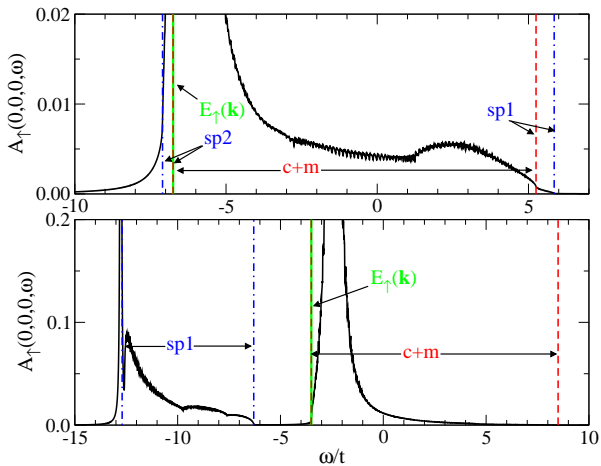}
  \caption{(color online) Spectral weight
    $A_{\uparrow}(\mathbf{k}=0,\omega)$ for 
    the 3D Ising  model at $\beta t =0.5$ for  FM $J_0/t=-3$ (top) and AFM
    $J_0/t=10$ (bottom) couplings. The edges of the c+m 
   continuum (dashed red lines) and sp1/sp2 continuum (dot-dashed blue
   lines) are indicated, as is $E_{\uparrow}(0)$ (thick green
   line). Other parameters are
    $J/t=0.5,  S=0.5, \eta=0.01$. } 
 \label{fig:3dI}
\end{figure}

So far we have done the whole analysis in 2D, simply because the
calculation of $\Sigma(\mathbf{k},\omega)$, especially for the
Heisenberg model, is numerically much faster. However, we did
investigate the 3D models and found essentially the same physics. As 
 examples, in Figs. \ref{fig:3dH}, \ref{fig:3dI} we show 
spectra for both FM $J_0/t=-3$ and AFM $J_0/t=10$, for both
models. These spectra display exactly the same features as the
corresponding 2D spectra. For the Heisenberg model we 
chose a larger $\eta=0.1$ and decreased the linear system size 
drastically to keep the computational time reasonable.
Consequently, the continuum edges are more difficult to discern, while 
finite size effects are more pronounced. In any event, the knowledge
accumulated from analysing the 2D data is fully consistent with all
features we observed in all 3D data we generated.

\section{Conclusions} \label{sec:conclusions}

To summarize, we calculated analytically the
lowest-$T$ correction to the self-energy of a spin-up carrier injected
in a FM background. We used both Heisenberg and Ising couplings to
describe the background,  to understand the relevance of gapped
vs. gapless magnons. These results  show how the
spectral weight evolves from  a discrete peak at $T=0$
to a collection of continua for $T \ne 0$ (these can merge, in the appropriate
circumstances), and explain their origin and how their locations
can be inferred. 

We were aided in this task by the fact that this model conserves the
$z$-component of the total spin, allowing us to consider the
contribution to the spectral weight coming from Hilbert subspaces with
different numbers $m$  of magnons  when the carrier has spin
up. Although we focused on the $m=1$, lowest-$T$ contribution, based on
the knowledge we acquired we can extrapolate with some confidence to
higher-$T$, as we discuss now.

One definite conclusion of this work is that knowledge of the $T=0$
carrier spectrum (in the $m=0$ sector) $E_{\uparrow}(\mathbf{k})$, and
of the magnon dispersion, $\Omega_{\mathbf{q}}$, is generally not
sufficient to predict a priori all features of the finite-$T$ spectral
weight, although a fair amount can be inferred from them. To see why, let us assume
that magnons do not interact with one another. (This is not true for
either model, for example due to their hard-core repulsion; we will
return to possible consequences of their interactions below.) If magnons
were non-interacting, then Lehmann decomposition of the higher-order
contributions in Eq. (\ref{eq:GF-time}) would predict finite-$T$
spectral weight for all intervals $\{
E^{(m)}_{\alpha}(\mathbf{k}+\sum_{i=1}^{m}\mathbf{q}_i)-
\sum_{i=1}^{m}\Omega_{\mathbf{q}_i}\}_{\mathbf{q}_1+\dots+\mathbf{q}_m}$,
$m=0, 1, \dots$.

Since we move from the $m$ to the $m+1$ subspace by adding a
magnon, and given that total momentum is conserved, we know that 
the spectrum in  subspace $m+1$ necessarily includes the convolution between
the spectrum of the subspace $m$ and the magnon dispersion, {\em
  i.e.}   $\{ E^{(m)}_{\alpha}(\mathbf{k}-\mathbf{q})+
\Omega_{\mathbf{q}}\}_{\mathbf{q}}$ is part of the spectrum  $
E^{(m+1)}_{\alpha}(\mathbf{k})$ (these are the scattering
states between the extra magnon and any eigenstate in the $m$
spectrum). 

This observation allows us to infer the
location of some of the finite-$T$ spectral weight, by recurrence. 
$E^{(1)}_{\alpha}(\mathbf{k})$ must include all scattering states $\{
E_{\uparrow}(\mathbf{k}-\mathbf{q})+
\Omega_{\mathbf{q}}\}_{\mathbf{q}}$, so the $m=1$ contribution
to the spectrum must span  $\{ E^{(1)}_{\alpha}(\mathbf{k}+\mathbf{q}')-
\Omega_{\mathbf{q}'}\}_{\mathbf{q}'} = \{ E_{\uparrow}(\mathbf{k}-\mathbf{q}+\mathbf{q}')-
\Omega_{\mathbf{q}'}+\Omega_{\mathbf{q}}\}_{{\mathbf{q}},{\mathbf{q}'}}$. We
called this the c+m continuum and verified that it is indeed 
seen in
the finite-$T$ spectral weight. Knowledge of this part of the $m=1$
spectrum allows us to infer scattering states that are part of the
$m=2$ spectrum and therefore their Lehmann contribution, etc.  The
conclusion is that all intervals $\{
E_{\uparrow}(\mathbf{k}+\sum_{i=1}^{m}\mathbf{q}'_i-\sum_{i=1}^{m}\mathbf{q}_i)- 
\sum_{i=1}^{m}\Omega_{\mathbf{q}'_i}+\sum_{i=1}^{m}\Omega_{\mathbf{q}_i}
\}_{{\mathbf{q}_1},\dots,{\mathbf{q}'_m}}$ will contain some spectral
weight at finite-$T$. For the dispersionless Ising  magnons 
this interval is the same for all $m$. For dispersive Heisenberg magnons this
interval broadens with $m$. For very small $J$, the additional
broadening as $m$ increases is very small and moreover one would
expect little spectral weight in the high-$m$ sectors if the
$T$ is not too large. Thus, we expect weight to be visible in
the c+m continuum up to high(er) temperatures; its boundaries may
also slowly expand with $T$, for a Heisenberg background, as
 higher $m$ subspaces become thermally activated.

Apart from these scattering states, $ E^{(m+1)}_{\alpha}(\mathbf{k})$ might
also contain bound states where the extra magnon is coherently bound
to all the other particles. The existence and location of such
coherent states cannot be predicted a priori, as they depend on the
details of the model (however, they certainly cannot appear unless
coherent states exist in the $m$ space).
An example  is the $E^{(1)}_{\alpha}(\mathbf{k})$ spectrum which
indeed contains the scattering states discussed above, but 
also contains the sp1 and/or sp2 discrete polarons states. These
give rise to their own continua  of scattering states
in higher $m$ subspaces, whose locations can be inferred by recurrence.

The question, then, is if it is likely to find such new, bound
coherent states for all values of $m$, i.e. if the number of
additional continua becomes arbitrarily large with increasing
$T$. Generally, the answer must be ``no'', since this requires bound
states between arbitrarily large numbers 
of objects. For the problem at hand, we believe that
it is quite unlikely that they appear even in the 
$m=2$ subspace, since that would involve one carrier binding two
magnons. This is a difficult task given the weak nearest-neighbour
attraction of order $J$ between magnons (due to the breaking of fewer
FM bonds), and the fact that the carrier can interact with only one
magnon at a time. The exception is likely to be in 1D systems where
magnons can coalesce into magnetic domains.

Let us now consider the role of magnon interactions. Because
of them, many-magnon states are not eigenstates of the Heisenberg
Hamiltonian so higher-order terms are not obtained by 
tracing over states with many independent magnons (in the Ising model
this complication can be avoided by working in real space). If the
attraction between magnons is too weak to bind them, this is
not an issue since their spectra will still consist of 
scattering states spanning the same energies like for non-interacting
magnons. As a result, the location of various features is not
affected, but the distribution of the spectral weight inside them will
be since the eigenfunctions are different. Magnon
pairing is unlikely for $d >1$  unless the exchange is
strongly anisotropic. However, if it happens and if the
spectrum of the magnon pairs is known, one could infer its effects
on the carrier spectral weight just like above.

Based on these arguments, we expect the higher-$T$ spectral
weight to show the same features we uncovered at low-$T$ (the
distribution of the weight between them might be quite
different, though). These expectations could be verified with
numerical simulations 
(conversely, our low-$T$ results can be used to test codes). Such
simulations would also solve the issue of the discrete 
peak that we observed for AFM $J_0$, and which we argued to be an artefact of our
low-$T$ approximation.

To conclude, although quantitatively our results are only  valid at
extremely low-$T$, we believe that this study clarifies qualitatively
how the spectral weight of a spin-up carrier evolves with $T$. Our
arguments can be straightforwardly extended to predict what features
appear in the spectral weight of a spin-down carrier, as well. 

 A general feature demonstrated by our work is that finite-$T$ does not
  result in just a simple thermal broadening of the quasiparticle peak,
 as it becomes a resonance inside a continuum.  Spectral weight can
 also be transferred to quite different energies if the quasiparticle can
 bind additional magnons into coherent polarons. When this happens,
 interpretation of experimentally measured and/or of computationally
 generated spectra could become
 difficult, unless one is aware of this possibility. 

\acknowledgements This work was supported by NSERC and QMI.

\appendix

\section{Derivation of the lowest $T\ne 0$ self-energy term}

We present this calculation for the Heisenberg FM; the Ising
case is treated similarly.
To find $G_{\uparrow}^{(1)}(\mathbf{k}, \mathbf{q},
\mathbf{q}',\omega)$, we divide $\hat{H}
=\hat{H}_0+\hat{V}$ where $\hat{H}_0=\hat{T}+\hat{H}_{\text{S}}^z$ and
$\hat{V}=\hat{H}_{\text{S}}^{x,y}+ \hat{H}_{\text{exc}} $, and use Dyson's identity
$\hat{G}(\omega)
=\hat{G}_0(\omega)+\hat{G}(\omega)\hat{V}\hat{G}_0(\omega)$ where
$\hat{G}_0=[\omega-\hat{H}_0+i \eta]^{-1}$ is the resolvent for
$\hat{H}_0$. This procedure is similar to that used in
Ref. \onlinecite{Berciu-sp} for the $T=0$ spin-polaron. Applying
Dyson's identity once we obtain: 
\begin{align}
& G_{\uparrow}^{(1)}(\mathbf{k}, \mathbf{q},
  \mathbf{q}',\omega)=G^{(0)}_{\uparrow}(\mathbf{k+q'-q},\omega+\Omega_{\mathbf{q'}}-\Omega_{\mathbf{q}})
  \left [ 
    \delta_{\mathbf{q},\mathbf{q'}}\right .  \nonumber \\
  &\left . -\frac{J_0}{2 N} \sum_{\mathbf{Q}} G_{\uparrow}^{(1)}(\mathbf{k}, \mathbf{Q}, \mathbf{q}',\omega) 
 + J_0\sqrt{\frac{S}{2N}} F(\mathbf{k},\mathbf{q},\omega)  \right]
 \label{eq:eomG}
\end{align}
The first term on the right-hand side is just the diagonal term. The
second term accounts for the energy shift that occurs when the
up-carrier is on the same site as the magnon, and the third term
contains a new propagator $F(\mathbf{k},\mathbf{q'},\omega)=
\langle
\Phi(\mathbf{q'})|c_{\mathbf{k},\uparrow} \hat{G}(\omega+\Omega_{q'})
c_{\mathbf{k+q},\downarrow}^\dagger |{\rm FM} \rangle$. This term accounts for
spin-flip processes where the up-carrier absorbs the magnon, turning
into a down-carrier with momentum  
$\mathbf{k+q}$. 
Using Dyson's identity again, we get an equation of motion for
$F(\mathbf{k},\mathbf{q'},\omega)$: 
\begin{align}
 F(\mathbf{k},\mathbf{q'},\omega)= &J_0 \sqrt{\frac{S}{2
     N}}G^{(0)}_{\uparrow}(\mathbf{k+q},\omega+\Omega_{\mathbf{q'}}+J_0
 S) \nonumber \\&\times \sum_{\mathbf{Q}}
 G_{\uparrow}^{(1)}(\mathbf{k}, \mathbf{Q}, \mathbf{q}',\omega).
 \label{eq:eomF}
\end{align}
The diagonal element vanishes
since the bra and ket are orthogonal. The energy shift
$-J_0S/2$ of the 
spin-down carrier is absorbed into the argument of
$G^{(0)}_{\uparrow}$, leaving only  the spin-flip process
which links $F$ back to $G_{\uparrow}^{(1)}$. 

These two coupled equations can now be solved as follows. We insert
Eq. (\ref{eq:eomF}) into Eq. (\ref{eq:eomG}) to obtain: 
\begin{align} 
& G_{\uparrow}^{(1)}(\mathbf{k}, \mathbf{q},
  \mathbf{q}',\omega)=G^{(0)}_{\uparrow}(\mathbf{k+q'-q},
  \omega+\Omega_{\mathbf{q'}}-\Omega_{\mathbf{q}})   
  \left \{  \delta_{\mathbf{q},\mathbf{q'}}\right .  \nonumber \\
  &\left . 
 -\frac{J_0}{2 } f(\mathbf{k},\mathbf{q'},\omega) \left [ 1- J_0 S
   G^{(0)}_{\uparrow}(\mathbf{k+q},\omega+\Omega_{\mathbf{q'}}+J_0 S)
   \right ]  \right \}, 
  \label{eq:eom-combined}
\end{align}
where $f(\mathbf{k},\mathbf{q'},\omega)=\frac{1}{N}
\sum_{\mathbf{Q}} G_{\uparrow}^{(1)}(\mathbf{k}, \mathbf{Q},
\mathbf{q'},\omega)$. Using Eq. (\ref{eq:eom-combined}) in the
definition of $f(\mathbf{k},\mathbf{q'},\omega)$ yields: 
\begin{align}
 &f(\mathbf{k},\mathbf{q'},\omega)=\frac{1}{N}
 G^{(0)}_{\uparrow}(\mathbf{k}, \omega)\left[1+\frac{J_0}{2}
   g(\mathbf{k},\mathbf{q'}, \omega) \right. \nonumber \\
&\left. \times \left(1- J_0 S
     G^{(0)}_{\uparrow}(\mathbf{k+q'},\omega+J_0 S)
     \right)\right]^{-1}, \nonumber  
\end{align}
 with
$ g(\mathbf{k},\mathbf{q'},\omega)=\frac{1}{N} \sum_{\mathbf{Q}}
 G^{(0)}_{\uparrow}(\mathbf{k+q'-Q},\omega+\Omega_{\mathbf{q'}}-
 \Omega_{\mathbf{Q}}).$ Note that $g(\mathbf{k},\mathbf{q'},\omega)$
 can be calculated numerically since $G^{(0)}_{\uparrow}(\mathbf{k},
 \omega)$ is a known function. 

All that is left to do is to insert
 the above expression into Eq. (\ref{eq:eom-combined}) and
 calculate  $ \sum_{\mathbf{q}}\exponential{-\beta
   \Omega_{\mathbf{q}}}  G_{\uparrow}^{(1)}(\mathbf{k}, \mathbf{q},
 \mathbf{q},\omega)$, to find the expression listed in Section III.


\begin{thebibliography}{10}
\bibitem{note} The
  prefactor contains the Riemann number $\zeta(1)=\infty$. However, this
  approximation ignores non-parabolic corrections to the dispersion
  and lacks the equivalent of a ``Debye 
  energy'',  instead summing over all energies. Including such corrections
  should remove the  
  singularity.
\end{thebibliography}
\end{document}